# Epistemic Impact on Group Problem Solving for Different Science Majors[*]

Andrew J. Mason, Charles A. Bertram

University of Central Arkansas, Conway, USA

Implementation of cognitive apprenticeship in an introductory physics lab group problem solving exercise may be mitigated by epistemic views toward physics of non-physics science majors. Quantitative pre-post data of the Force Concept Inventory (FCI) and Colorado Learning Attitudes About Science Survey (CLASS) of 39 students of a first-semester algebra-based introductory physics course, while describing typical results for a traditional-format course overall ($g = +0.14$), suggest differences in epistemic views between health science majors and life science majors which may correlate with differences in pre-post conceptual understanding. Audiovisual data of student lab groups working on a context-rich problem and students' written reflections described each group's typical dynamics and invoked epistemic games. We examined the effects of framework-based orientation (favored by biology majors) and performance-based orientation (favored by computer science, chemistry, and health science majors) on pre-post attitude survey performance. We also investigated possible correlations of these orientations with individual quantitative survey results, and with qualitative audiovisual data of lab groups' choice of epistemic games.

## Introduction

Recently, attention has been turned toward designing Introductory Physics for the Life Sciences (IPLS) courses to suit the needs of life science majors (Redish et al., 2014; Moore, Giannini, & Losert, 2014). The importance of problem-solving skills for IPLS students is one of several concerns addressed by these efforts (Crouch & Heller, 2014). Techniques, such as coordinated group problem-solving (Heller, Keith, & Anderson, 1992), may be a useful way to introduce students to physics problem-solving techniques, in conjunction with context-rich problems (Heller & Hollabaugh, 1992) designed to assist students in developing a problem-solving framework.

One concern when addressing these issues regards measuring attitudes toward problem-solving, e.g., the Maryland Physics Expectations Test (MPEX) (Redish, Steinberg, & Saul, 1998) and the Colorado Learning Attitudes About Science Survey (CLASS) (Adams, Perkins, Podolefsky, Dubson, Finkelstein, & Wieman, 2006). In particular, it is noted that learning orientation of physical science majors often leads to more

[*] **Acknowledgements:** The authors of this paper are grateful to L. Ratz for his assistance in data collection. They are also grateful to L. Kryjevskaia, A. Boudreaux, A. McInerny, B. Lunk, and M. B. Kustusch for their helpful discussions. Funding was provided by the University of Central Arkansas Sponsored Programs Office and Department of Physics and Astronomy.

Andrew J. Mason, Ph.D., assistant professor, Department of Physics and Astronomy, University of Central Arkansas.

Charles A. Bertram, undergraduate student researcher, Department of Physics and Astronomy, University of Central Arkansas.

successful careers in the physical sciences (Hazari, Potvin, Tai, & Almarode, 2010); however, it remains unclear whether such a learning orientation exists toward physics for IPLS students who are not majoring in physics.

**An IPLS Course With Diversity of Majors: Are Expectations Different?**

Currently, we are investigating an algebra-based IPLS course taught at a medium-size state university level with moderate-sized student populations per course section (45-70 students). The format of the course is traditional in nature, and is taught by two to four instructors each semester in a traditional lecture-lab format, albeit with no recitation sections. A typical class population contains biology majors as well as chemistry and computer science majors, in addition to health and behavioral science majors who are housed in a separate college from the other majors. As a result, content importance, as well as course expectations, may fluctuate between different majors.

With regard to problem-solving, such a variance in expectations may possibly affect the learning outcomes and effectiveness of the exercise. One potential effect is lab groups' choice of epistemic games (Tuminaro & Redish, 2007) to approach the problem solution; performance-motivated students may choose games that seem to more directly reach the solution, ignoring other potentially useful games oriented toward a problem-solving framework.

We examined data from a typical first semester IPLS course section to investigate the nature of expectation differences and how they may affect student interactions in a group problem-solving exercise. A reflection exercise (Yerushalmi, Cohen, Mason, & Singh, 2012) is introduced to help students identify areas of struggle in solving physics problems. Epistemic and attitudinal tendencies among majors will be identified and considered in light of students' views toward a problem-solving framework.

**Research Goals**

We establish a preliminary measure of students' learning goals, and investigate whether this measure is related to survey results reflecting content knowledge and attitudes toward physics for the problem-solving exercise. Choice of major may be correlated to learning goals as well. We also examine whether this same measure of learning goals for laboratory groups corresponds to choice of epistemic games for a problem-solving approach.

## Procedure

**Lab Problem-Solving Exercise**

Data were taken in the Spring Semester of 2014 for a first semester introductory algebra-based physics course of 48 students spread across two laboratory sections. The study took place with primarily biological science and health and behavioral sciences majors; the student body also included several computer science and chemistry majors and some non-science majors. The problem-solving activity comprised the first hour of each lab section prior to the lab activity.

Beginning each laboratory period, a context-rich problem was introduced to lab groups of two or three students each to work on cooperatively. Students wrote and submitted their reflections about which part of the problem-solving process they individually struggled with, if any, on a rubric adapted from a self-diagnosis study done by Yerushalmi et al. (2012). A learning assistant (Otero, Pollock, & Finkelstein, 2010) was available to proctor the laboratory sections alongside the instructor. The instructor and the learning assistant assisted students in a verbal tutorial fashion; in addition, lab groups were permitted to use their lecture notes if desired. After the students were finished, the instructor outlined the solution and allowed the students a chance to reflect on their problem-solving skills, specifically the areas of the problem solution that they struggled to understand along the way. This process was repeated every week for the duration of the semester, except exam weeks.

**Data Collection**

The Force Concept Inventory (FCI) and the CLASS were given as pre- and post- tests on the first and last laboratory sections of the semester, with emphasis on the CLASS for attitudinal changes in student population, both overall and with regard to specific item clusters. Of the two laboratory sections, 39 total students submitted complete data for both surveys. Four students, two from each lab section, were omitted for either failing to provide complete data or not taking a form of data seriously (e.g., choosing "Neutral" for all CLASS questions). Five other students dropped the course prior to its conclusion.

Audiovisual data were taken of the students on a late-semester laboratory lesson regarding rotational dynamics. The researchers used these data to confirm typical lab group behavior observed anecdotally over the course of the semester, and to identify epistemic games used by different lab groups in the transcribed data. The latter goal is useful to understand whether or not choice of epistemic games is related to FCI gains, CLASS gains, or student survey responses.

In addition, the students were given an end-of-semester survey to provide feedback in free-response form about the reflection exercise. The survey asked the question: "In what ways did you find the exercise useful toward learning the material in the course?". The students' written responses were collected and transcribed in order to determine a classification scheme.

## Results

**Free-Response Survey: Measure of Learning Goals**

Responses of the students on the end-of-semester survey showed that 37 of 39 students found the learning exercise useful as a whole. The remaining two students left comments about the portions of the exercise they did find useful, and so could still be analyzed. After transcription, the survey responses were found to be classifiable into one of the three groups.

The first response group was more "framework-oriented", i.e., responses focused on receiving help on different aspects of a problem-solving framework (e.g., visualization, concepts, and mapping to equations). The second group was "performance-oriented", i.e., their responses were focused on how the exercise helped them perform on other aspects of the course (e.g., exams, homework, and pre-laboratory preparation). The remainder of students' responses offered responses that could not be definitively classified, either because the response was too general (e.g., helping on problem-solving as a whole) or because the response did not seem to focus strongly on specific aspects of the course material (e.g., studying in a group). As such, all 39 students were classifiable by survey responses.

Table 1 shows the distribution of lab sections, as well as the distribution of select majors into response categories, namely, "framework-oriented", "performance-oriented", and "vaguely-defined" response categories. It is noted that three students gave both framework-oriented and performance-oriented responses, and so are included in both of those categories.

Table 1

*Students Categorized Into Problem-Solving Exercise Orientations as Determined by End-of-Semester Free-Response Essays*

| Group | Framework-oriented | Performance-oriented | Vaguely-defined |
|---|---|---|---|
| $N$ (out of 39) | 14 | 19 | 9 |
| Lab 1 (19) | 8 | 9 | 5 |
| Lab 2 (20) | 6 | 10 | 4 |
| Biological science (16) | 10 | 5 | 3 |
| Health and behavioral sciences (10) | 2 | 5 | 4 |
| Chemistry/computer science (8) | 0 | 8 | 0 |
| Non-science (5) | 2 | 1 | 2 |

To properly evaluate whether these categories are meaningful, we consider potential overlap with the FCI and the CLASS. Table 2 compares the average pre-test scores and average individual normalized gains for students on the FCI and CLASS with regard to the three student categories defined in Table 1. FCI data are in terms of percentages of correct responses, and CLASS data are in terms of percentages of responses that are considered more "expert-like", as opposed to neutral or more novice-like responses. As there are individual fluctuations in pre-tests and post-tests, an average of individual student gains is different, and more accurate, than comparing the group pre-test score to the group post-test score; hence, post-test scores are omitted to avoid confusion.

Table 2

*Average Scores in Percentile Form for Student Survey Response Groups for FCI and CLASS Pre-test Scores, as Well as Averaged Individual Modified Gains on the Post-test for both the FCI and CLASS*

| Group | Framework-oriented | Performance-oriented | Vaguely-defined | All |
|---|---|---|---|---|
| $N$ | 14 | 19 | 9 | 39 |
| FCI pre-test (%) | 28% | 31% | 24% | 30% |
| FCI gain ($g$) | +0.22 | +0.13 | +0.04 | +0.14 |
| CLASS pre-test (%) | 62% | 62% | 52% | 61% |
| CLASS gain | +0.10 | -0.09 | +0.00 | -0.02 |

*Note. SE* ranges from ±2-7% for FCI data and ±4-9% for CLASS data.

The framework-oriented response group appeared to have the highest averaged individual gain of the three groups. The vaguely-defined response group had minimal FCI gains and negligible CLASS gains, while the performance-oriented group experienced moderate FCI gains and an averaged decline across individual CLASS gains.

With regard to CLASS item clusters, the framework-oriented group experienced moderately strong positive gains for the three problem-solving item clusters ($g = +0.34$ for the PS-General cluster, +0.46 for PS-Confidence, and +0.16 for PS-Sophistication), as well as weak positive gains for Conceptual Understanding and Applied Conceptual Understanding clusters ($g = +0.08$). In contrast, the performance-oriented group had negative gains across all item clusters, with gain values ranging from -0.09 to -0.32.

Despite small sample size, a borderline significant difference existed in average gains of framework-oriented and performance-oriented students ($p = 0.06$). The framework-oriented group also showed borderline significance in FCI gains from the vaguely-defined group ($p = 0.056$). There were no other statistically significant differences between the three groups.

PROBLEM SOLVING FOR DIFFERENT SCIENCE MAJORS                                                    5**Use of Epistemic Games**

Transcriptions from the audiovisual data sample provided examples of epistemic games that could be corroborated with corresponding reflection rubrics submitted for that particular lab problem. Figure 1 displays the rubric reflections of a student that was classified as framework-oriented. The student here invokes these games in terms of reflection upon areas of the problem solution which presented a struggle.

|  | Performance Level | What did you struggle with? |
|---|---|---|
| **Problem description** sum of torques using sum of inertia times angular acceleration | (Full)/Partial/Missing | Understanding $T = F\ell$ is not the desired formula, instead $T = I\alpha$ is best fit |
| **Solution construction** $\Sigma T = (\Sigma I)\alpha$ | (Full)/Partial/Missing | |
| **Logical progression** The weight of the hanging mass is irrelevant if we use the $a_T$ given | Full/(Partial)/Missing | neglecting force of tension |

*Figure 1.* Example of a framework-based student's rubric from a lab group problem-solving exercise involving rotational dynamics.

In terms of games identified by Tuminaro and Redish (2007), the student exhibits an example of a Recursive Plug and Chug game, i.e., the student realized one equation for torque does not address the unknown quantity of angular acceleration, and opted for another equation that does. The student also hints at use of a Mapping Mathematics to Meaning game; the written phrase "The weight of the hanging mass is irrelevant if we use the $a_T$ given" implies recognition of a target concept, namely, the relationship between tangential linear acceleration and angular acceleration, which tells the story of why a force from a hanging mass does not need to be calculated.

Audiovisual data, as checked by rubric performance by the researchers, showed an interesting distribution of epistemic games preferred by different lab groups, as described by Tuminaro and Redish (2007). The results are shown in Table 3, in terms of lab groups that were majority (at least two members who were) performance-oriented, majority framework-oriented, or an even mix of performance and framework-oriented response students. One group with two vaguely-defined response students is omitted, as this group hesitated to use any epistemic games at all. The Physical Mechanism game was present for all lab groups, as the context-rich problem explicitly prompted its use with constructing a story about the physical situation; as such, the game is not included in Table 3. Lab groups tended to have either two or three performance-oriented students (performance groups) or a mixture of performance-oriented, framework-oriented, and vaguely-defined response students (mixed groups). Performance and mixed groups both heavily used the Recursive Plug and Chug game, while performance groups also focused strongly on Mapping Math(ematics) to Meaning in trying to find an equation that would relate the target to other problem concepts. However, this game cannot explicitly be tied to performance-oriented students, as several performance groups had a non-performance-oriented student with two performance-oriented students.

Table 3

*Tendencies of Different Orientations of Lab Groups Using Given Epistemic Games*

| Game | Performance groups | Framework groups | Mixed groups | Total |
| --- | --- | --- | --- | --- |
| Pictorial Analysis | 3 | 2 | 1 | 6 |
| Transliteration to Maths | 1 | 2 | 1 | 4 |
| Recursive Plug and Chug | 5 | 0 | 5 | 10 |
| Mapping Math to Meaning | 5 | 1 | 1 | 7 |
| Mapping Meaning to Math | 0 | 0 | 0 | 0 |
| Total Lab Groups | 6 | 2 | 6 | 14 |

*Note.* Physical Mechanism is omitted.

## Discussion and Conclusion

Framework-oriented student survey responses are suggested by CLASS and FCI results to translate to higher relative individual gains in conceptual understanding and higher relative individual gains in attitudes toward physics. Performance-oriented student survey responses translate to somewhat less robust FCI gains and broadly negative CLASS gains. Of interest is an indication from Table 1 that most framework-oriented students were biology majors, accompanied by two health science majors and two non-science majors. Taken in context with the results in Table 2, biology majors seem more likely to employ a framework-oriented perspective and benefit more in content and attitudinal gains.

Performance-oriented lab groups heavily, but not exclusively, favor a choice of epistemic games, which, as observed, seem to reflect a heavy reliance on using equations and typify a novice-like problem-solving approach. These tendencies seem to support CLASS data suggesting a decline in expert-like attitudes for performance-oriented students. While sample size is low for framework-oriented groups, there is a tendency to focus on Transliteration to Mathematics and Pictorial Analysis games, which correspond to a wider variety of problem-solving framework elements.

While biology majors were more likely to prefer and benefit from a framework-oriented view, this is not true for chemistry and computer science majors, who prefer performance, or for HBS majors, who either prefer performance or have a vaguely-defined view of problem-solving. Thus far, diversity of majors appears to be linked to an epistemic disparity that presents a challenge to problem-solving framework pedagogies. Future class sections will offer a larger sample size in order to further define these tendencies.